\documentclass[a4paper,11pt]{article}
\usepackage{pos}

\usepackage{textcomp}
\usepackage{amsfonts} 
\usepackage{amsmath} 
\usepackage{slashed}
\usepackage{pifont}
\usepackage{multirow,lscape}
\usepackage{array}
\usepackage{subcaption}
\usepackage{verbatim}
\usepackage{bm}
\usepackage{comment}
\usepackage{xcolor}
\usepackage{url} 

\graphicspath{{./fig/}}

\providecommand{\wbar}[1]{\overline#1}

\providecommand{\mate}[3]{\langle#1\lvert#2\rvert#3\rangle}

\renewcommand{\Re}{\mathrm{Re}\,}
\renewcommand{\Im}{\mathrm{Im}\,}

\providecommand{\GeV}{\;\mathrm{GeV}}

\definecolor{HLBlue}{HTML}{6599FF}
\definecolor{HLOrange}{HTML}{FF6600}

\newcommand{\BK}{\hat{B}_{K}}
\newcommand{\Vcb}{|V_{cb}|}

\newcommand{\Vus}{|V_{us}|}
\newcommand{\Vud}{|V_{ud}|}

\newcommand{\eps}{\varepsilon}

\newcommand{\epsK}{\varepsilon_{K}}

\newcommand{\BtoDstp}{\bar{B} \to D^{(\ast)} \ell \bar{\nu}}
\newcommand{\BtoDst}{\bar{B} \to D^\ast \ell \bar{\nu}}

\newcommand{\red}[1]{\textcolor{red}{#1}} 
\newcommand{\blue}[1]{\textcolor{blue}{#1}}

\begin{document}
\title{2023 Update of $\epsK$ with lattice QCD inputs}
\ShortTitle{$\epsK$ with lattice QCD inputs}

\author[a]{Seungyeob Jwa}
\author[a]{Jeehun Kim}
\author[a]{Sunghee Kim}
\author[a]{Sunkyu Lee}
\author*[a,b,1]{Weonjong Lee}
\author[c]{Jaehoon Leem}
\author[a]{Jeonghwan Pak}
\author[d]{Sungwoo Park}

\affiliation[a]{Lattice Gauge Theory Research Center, CTP, and FPRD,
  Department of Physics and Astronomy, \\
  Seoul National University,
  Seoul 08826, South Korea}


\affiliation[b]{School of Physics,
  Korea Institute for Advanced Study (KIAS),
  Seoul 02455, South Korea}

\affiliation[c]{Computational Science and Engineering Team,
  Innovation Center, Samsung Electronics, Hwaseong,
  Gyeonggi-do 18448, South Korea.}

\affiliation[d]{Lawrence Livermore National Lab, 7000 East Ave,
      Livermore, CA 94550, USA}

\emailAdd{wlee@snu.ac.kr}

\note{For the SWME collaboration}

\abstract{
  We report recent progress on $\epsK$ evaluated directly from the
  standard model (SM) with lattice QCD inputs such as $\BK$, $\Vcb$,
  $\Vus$, $\Vud$, $\xi_0$, $\xi_2$, $\xi_\text{LD}$, $f_K$, and $m_c$.
  We find that the standard model with exclusive $\Vcb$ and lattice
  QCD inputs describes only 66\% of the experimental value of
  $|\epsK|$ and does not explain its remaining 34\%, which corresponds
  to a strong tension in $|\epsK|$ at the $4.9\sigma \sim 3.9\sigma$
  level between the SM theory and experiment. We also find that this
  tension disappears when we use the inclusive value of $\Vcb$
  obtained using the heavy quark expansion based on the QCD sum rule
  approach. }

\FullConference{The 40th International Symposium on Lattice Field
  Theory (Lattice 2023)\\ July 31st - August 4th, 2023\\ Fermi
  National Accelerator Laboratory\\}

\maketitle

%
%
\section{Introduction}
This paper is an update of our previous reports \cite{ Lee:2023lxz,
  Lee:2021crz, Kim:2019vic, Bailey:2018feb, Bailey:2015tba,
  Bailey:2018aks, Jang:2017ieg, Bailey:2015frw}.
We report recent progress in the determination of $|\epsK|$ with
updated inputs from lattice QCD.
Updated input parameters include $\lambda$, $\bar{\rho}$,
$\bar{\eta}$, exclusive $\Vcb$, $\Vus$, $\Vud$, $\Vus/\Vud$, $M_W$, and
$M_t$.

We follow the color convention of our previous papers \cite{
  Lee:2023lxz, Lee:2021crz, Kim:2019vic, Bailey:2018feb,
  Bailey:2015tba, Bailey:2018aks, Jang:2017ieg, Bailey:2015frw} in
Tables \ref{tab:input-Vus-Vud}--\ref{tab:epsK}.
We use red for new input data used to evaluate $\epsK$.
We use blue for new input data which is not used.


%
%
%
\section{Input parameters: Wolfenstein parameters}
\label{sec:wp}
In Table \ref{tab:input-Vus-Vud}, we summarize results for $\Vud$,
$\Vus$ from lattice QCD and nuclear $\beta$ decays, and
$\dfrac{\Vus}{\Vud}$ from lattice QCD. 
\begin{table}[h!]
  \begin{subtable}{0.60\linewidth}
    \renewcommand{\arraystretch}{1.2}
    \centering
    \resizebox{1.0\linewidth}{!}{
      \begin{tabular}{ l @{\quad} l @{\quad} l @{\quad} c }
        \hline\hline
        type  &  $\Vus$  & $\Vud$  & Ref.
        \\ \hline
        Lattice $N_f = 2+1+1$ & 0.2248(6)       & 0.97440(15)
        & FLAG-23 \cite{FlavourLatticeAveragingGroupFLAG:2021npn}
        \\
        Lattice $N_f = 2+1$   & \blue{0.2249(5)} & \blue{0.97438(12)}
        & FLAG-23 \cite{FlavourLatticeAveragingGroupFLAG:2021npn}
        \\ \hline
        nuclear $\beta$ decay & 0.2278(6)       & 0.97370(14)
        & PDG-21 \cite{Zyla:2020zbs}
        \\ \hline\hline
      \end{tabular}
    } 
    \caption{$\Vus$ and $\Vud$ from lattice QCD and nuclear $\beta$ decays}
    \label{tab:Vus+Vud}
  \end{subtable} 
  \hfill
  \begin{subtable}{0.38\linewidth}
    \renewcommand{\arraystretch}{1.65}
    \centering
    \resizebox{1.0\linewidth}{!}{
      \begin{tabular}{ l @{\quad} l @{\quad} c }
        \hline\hline
        type  &  $\Vus/\Vud$  & Ref.
        \\ \hline
        QCD     & 0.2313(5) 
        & FLAG-23 \cite{FlavourLatticeAveragingGroupFLAG:2021npn}
        \\
        QCD+QED & \red{0.2320(5)}
        & FLAG-23 \cite{FlavourLatticeAveragingGroupFLAG:2021npn}
        \\ \hline\hline
      \end{tabular}
    } 
    \caption{$\Vus/\Vud$ from lattice QCD}
    \label{tab:Vus/Vud}
  \end{subtable} 
  \caption{ (\subref{tab:Vus+Vud}) $\Vus$ and $\Vud$
    (\subref{tab:Vus/Vud}) $\Vus/\Vud$. }
  \label{tab:input-Vus-Vud}
\end{table}

\begin{align}
  \lambda &= \frac{ \Vus }{ \sqrt{ \Vud^2 + \Vus^2 } }
  = \frac{ r }{ \sqrt{ 1 + r^2 } }
  \,, \qquad\qquad
  r = \frac{\Vus}{\Vud}
  \label{eq:lambda-1}
\end{align}
Using Eq.~\eqref{eq:lambda-1}, we determine $\lambda$ from
$\dfrac{\Vus}{\Vud}$ in Table \ref{tab:input-Vus-Vud}
(\subref{tab:Vus/Vud}), because its error is less than that obtained
by using the values for $\Vus$ and $\Vud$ in Table
\ref{tab:input-Vus-Vud} (\subref{tab:Vus+Vud}).
Results for $\lambda$ are presented in Table
\ref{tab:input-WP-eta}\;(\subref{tab:WP}), where we summarize the most
updated Wolfenstein parameters (WP).

Recently the UTfit collaboration updated values for the WP in
Ref.~\cite{ UTfit:2022hsi}.
As explained in Ref.~\cite{ Bailey:2018feb, Bailey:2015frw}, we use
the results of the angle-only-fit (AOF) in Table
\ref{tab:input-WP-eta} (\subref{tab:WP}) in order to avoid unwanted
correlations between $(\epsK, \Vcb)$, and $(\bar\rho, \bar\eta)$.
We determine the parameter $A$ from $\Vcb$.
\begin{table}[h!]
  \begin{subtable}{0.73\linewidth}
    \renewcommand{\arraystretch}{1.2}
    \resizebox{1.0\linewidth}{!}{
      \begin{tabular}{ @{\qquad} c @{\qquad} | l l | l l | l l }
        \hline\hline
        WP
        & \multicolumn{2}{c|}{CKMfitter}
        & \multicolumn{2}{c|}{UTfit}
        & \multicolumn{2}{c}{AOF}
        \\ \hline
        $\lambda$
        & 0.22500(24)         & \cite{Charles:2004jd}
        & \blue{0.22519(83)}  & \cite{UTfit:2022hsi, Bona:2006ah}
        & \red{0.22600(46)}   & \cite{FlavourLatticeAveragingGroupFLAG:2021npn}
        \\ \hline
        $\bar{\rho}$
        & $0.1566(85)$        & \cite{Charles:2004jd}
        & \blue{$0.161(10)$}  & \cite{UTfit:2022hsi, Bona:2006ah}
        & \red{ $0.156(17)$ } & \cite{UTfit:2022hsi}
        \\ \hline
        $\bar{\eta}$
        & $0.3475(118)$       & \cite{Charles:2004jd}
        & \blue{$0.347(10)$}  & \cite{UTfit:2022hsi, Bona:2006ah}
        & \red{ $0.334(12)$ } & \cite{UTfit:2022hsi}
        \\ \hline\hline
      \end{tabular}
    } 
    \caption{ Wolfenstein parameters}
    \label{tab:WP}
  \end{subtable} 
  \hfill
  \begin{subtable}{0.25\linewidth}
    \renewcommand{\arraystretch}{1.3}
    \resizebox{1.0\linewidth}{!}{
      \begin{tabular}[b]{ c l c }
        \hline\hline
        Input & Value & Ref.
        \\ \hline
        $\eta_{cc}$ & $1.72(27)$   & \cite{Bailey:2015tba}
        \\
        $\eta_{tt}$ & $0.5765(65)$ & \cite{Buras2008:PhysRevD.78.033005}
        \\
        $\eta_{ct}$ & $0.496(47)$  & \cite{Brod2010:prd.82.094026}
        \\ \hline\hline
      \end{tabular}
    } 
    \caption{$\eta_{ij}$}
    \label{tab:eta}
  \end{subtable} 
  \caption{ (\subref{tab:WP}) Wolfenstein parameters and
    (\subref{tab:eta}) QCD corrections: $\eta_{ij}$ with $i,j = c,t$.}
  \label{tab:input-WP-eta}
\end{table}

%

%
%

%
%
\section{Input parameters: $\Vcb$}
\label{sec:Vcb}
In Table \ref{tab:Vcb}, we summarize recent updated results for
exclusive $\Vcb$ and inclusive $\Vcb$.
In Table \ref{tab:Vcb} (\subref{tab:ex-Vcb}), we present updated
results for exclusive $\Vcb$ obtained by various groups: FNAL/MILC,
FLAG, HFLAV, and HPQCD.
They are consistent with one another within $1.0\sigma$ uncertainty.

In Table \ref{tab:Vcb} (\subref{tab:in-Vcb}), we present recent updated
results for inclusive $\Vcb$.
There are a number of attempts to determine inclusive $\Vcb$ from
lattice QCD, but all of them at present belong to the category of
exploratory study rather than that of precision calculation \cite{
  Barone:2022gkn}.
\begin{table}[t!]

  \begin{subtable}{1.0\linewidth}
    \renewcommand{\arraystretch}{1.2}
    \center
    \vspace*{-5mm}
    \resizebox{1.0\textwidth}{!}{
      \begin{tabular}{@{\qquad} l @{\qquad} l @{\qquad} l @{\qquad} l @{\qquad} l @{\qquad}}
        \hline\hline
        channel & value & method & ref & source \\ \hline
        $B\to D^* \ell \bar{\nu}$
        & \red{38.40(78)} & BGL
        & \cite{ FermilabLattice:2021cdg} p27e76 & FNAL/MILC-22
        \\ \hline
        ex-comb
        & \red{39.48(67)} & comb
        & \cite{ FlavourLatticeAveragingGroupFLAG:2021npn} p195e314 & FLAG-23
        \\
        ex-comb & \red{39.10(50)} & comb
        & \cite{ HeavyFlavorAveragingGroup:2022wzx} p120e221 & HFLAV-23  
        \\
        ex-comb & \red{39.31(54)(51)} & comb
        & \cite{ Harrison:2023dzh} p22e51 & HPQCD-23  
        \\ \hline\hline
      \end{tabular}
    } 
    \caption{Exclusive $\Vcb$ in units of $10^{-3}$.}
    \label{tab:ex-Vcb}
  \end{subtable}  
  \begin{subtable}{1.0\linewidth}
    \renewcommand{\arraystretch}{1.2}
    \center
    \vspace*{+2mm}
    \resizebox{1.0\textwidth}{!}{
      \begin{tabular}{ @{\qquad} l @{\qquad\qquad} l @{\qquad\qquad\qquad} l @{\qquad\qquad} l @{\qquad} }
        \hline\hline
        scheme        & value         & ref  & source \\ \hline
        kinetic scheme & \blue{42.16(51)}
        & \cite{ Bordone:2021oof} p1 & Gambino-21
        \\
        1S scheme      & \red{41.98(45)}
        & \cite{ HeavyFlavorAveragingGroup:2022wzx} p108e200 & HFLAV-23 
        \\ \hline\hline
      \end{tabular}
    } 
    \caption{Inclusive $\Vcb$ in units of $10^{-3}$.}
    \label{tab:in-Vcb}
  \end{subtable}
  \caption{ Results for (\subref{tab:ex-Vcb}) exclusive $\Vcb$ and
    (\subref{tab:in-Vcb}) inclusive $\Vcb$. The abbreviation p27e76 means
    Eq.~(76) on page 27.}
  \label{tab:Vcb}
\end{table}
%
%
%

%

%
%

%
%
\section{Input parameter $\xi_0$}
The absorptive part of long distance effects on $\epsK$ is parametrized
by $\xi_0$.
\begin{align}
  \xi_0  &= \frac{\Im A_0}{\Re A_0}, \qquad
  \xi_2 = \frac{\Im A_2}{\Re A_2}, \qquad
  \Re \left(\frac{\eps'}{\eps} \right) =
  \frac{\omega}{\sqrt{2} |\eps_K|} (\xi_2 - \xi_0) \,.
  \label{eq:e'/e:xi0}
\end{align}
There are two independent methods to determine $\xi_0$ in lattice QCD:
the indirect and direct methods.
The indirect method is to determine $\xi_0$ using
Eq.~\eqref{eq:e'/e:xi0} with lattice QCD results for $\xi_2$ combined
with experimental results for $\eps'/\eps$, $\epsK$, and $\omega$.
The direct method is to determine $\xi_0$ using the lattice QCD
results for $\Im A_0$, combined with experimental results for $\Re
A_0$.

In Table~\ref{tab:xi0-sum} (\subref{tab:exp-ReA0-ReA2-1}), we summarize
experimental results for $\Re A_0$ and $\Re A_2$.
In Table~\ref{tab:xi0-sum} (\subref{tab:ImA0-ImA2-1}), we summarize
lattice results for $\Im A_0$ and $\Im A_2$ calculated by RBC-UKQCD.
In Table~\ref{tab:xi0-sum} (\subref{tab:xi0-1}), we present results
for $\xi_0$ obtained by using the results in Table~\ref{tab:xi0-sum}
(\subref{tab:exp-ReA0-ReA2-1}) and (\subref{tab:ImA0-ImA2-1}).

Here we use the results of the indirect method for $\xi_0$ to evaluate
$\epsK$, since the systematic and statistical errors are much smaller
than those of the direct method.

\begin{table}[htbp]
  \begin{subtable}{1.0\linewidth}
    \renewcommand{\arraystretch}{1.2}
    \center
    \vspace*{-7mm}
    \resizebox{1.0\textwidth}{!}{
      \begin{tabular}{ @{\qquad} l @{\qquad} l @{\qquad\qquad} l @{\qquad\qquad} l @{\qquad} l @{\qquad\qquad} }
        \hline\hline
        parameter & method & value & Ref. & source \\ \hline
        $\Re A_0$ & exp & $3.3201(18) \times 10^{-7} \GeV$ &
        \cite{ Blum:2015ywa, Bai:2015nea}  & NA
        \\
        $\Re A_2$ & exp & $1.4787(31) \times 10^{-8} \GeV$ &
        \cite{ Blum:2015ywa} & NA
        \\ \hline
        $\omega$ & exp & $0.04454(12)$ &
        \cite{ Blum:2015ywa} & NA
        \\ \hline
        $|\epsK|$ & exp & $2.228(11) \times 10^{-3}$ &
        \cite{ Zyla:2020zbs} & PDG-2021
        \\
        $\Re(\eps'/\eps)$ & exp & $1.66(23) \times 10^{-3}$ &
        \cite{ Zyla:2020zbs} & PDG-2021
        \\ \hline\hline
      \end{tabular}
    } 
    \caption{Experimental results for $\omega$, $\Re A_0$ and $\Re A_2$.}
    \label{tab:exp-ReA0-ReA2-1}
  \end{subtable}
  \vspace*{4mm}
  \begin{subtable}{1.0\linewidth}
    \renewcommand{\arraystretch}{1.2}
    \center
    \resizebox{1.0\textwidth}{!}{
      \begin{tabular}{ @{\qquad} l @{\qquad} l @{\qquad}@{\qquad} l @{\qquad}@{\qquad} l @{\qquad} l }
        \hline\hline
        parameter & method & value ($\GeV$) & Ref. & source \\ \hline
        $\Im A_0$ & lattice & $-6.98(62)(144) \times 10^{-11}$ &
        \cite{ RBC:2020kdj} p4t1   & RBC-UK-2020 
        \\
        $\Im A_2$ & lattice & $-8.34(103) \times 10^{-13}$  &
        \cite{ RBC:2020kdj} p31e90 & RBC-UK-2020
        \\ \hline\hline
      \end{tabular}
    } 
    \caption{Results for $\Im A_0$, and $\Im A_2$ in lattice QCD. }
    \label{tab:ImA0-ImA2-1}
  \end{subtable}
  \vspace*{3mm}
  \begin{subtable}{1.0\linewidth}
    \renewcommand{\arraystretch}{1.2}
    \center
    \resizebox{1.0\textwidth}{!}{
      \begin{tabular}{@{\qquad} l @{\qquad\qquad} l @{\qquad\qquad} l @{\qquad\qquad} l @{\qquad\qquad} l @{\qquad} }
        \hline\hline
        parameter & method & value & ref & source \\ \hline
        $\xi_0$ & indirect & \red{ $-1.738(177) \times 10^{-4}$ }
        & \cite{ RBC:2020kdj} & SWME \\
        $\xi_0$ & direct  & $-2.102(472) \times 10^{-4}$
        & \cite{ RBC:2020kdj} & SWME \\ \hline\hline
      \end{tabular}
    } 
    \caption{ Results for $\xi_0$ obtained using the direct and indirect
      methods in lattice QCD. }
    \label{tab:xi0-1}
  \end{subtable}
  \caption{Results for $\xi_0$. Here, we use the same notation as in
    Table \ref{tab:Vcb}. The abbreviation p4t1 means Table 1 on page
    4.}
  \label{tab:xi0-sum}
\end{table}

%

%
%
\section{Input parameters: $\BK$,~ $\xi_\text{LD}$,~ and others}
The Flavour Lattice Averaging Group (FLAG) \cite{
  FlavourLatticeAveragingGroupFLAG:2021npn} reports results for $\BK$
in lattice QCD with $N_f=2$, $N_f=2+1$, and $N_f = 2+1+1$.
Here we use the result for $\BK$ with $N_f=2+1$, which is obtained
by taking an average over the four data points from BMW 11, Laiho 11,
RBC-UKQCD 14, and SWME 15 in Table
\ref{tab:input-BK-other}\;(\subref{tab:BK}).

\begin{table}[b!]
  \begin{subtable}{0.40\linewidth}
    \renewcommand{\arraystretch}{1.2}
    \resizebox{1.0\linewidth}{!}{
      \begin{tabular}{ l  l  l }
        \hline\hline
        Collaboration & Ref. & $\BK$  \\ \hline
        SWME 15       & \cite{Jang:2015sla} & $0.735(5)(36)$     \\
        RBC/UKQCD 14  & \cite{Blum:2014tka} & $0.7499(24)(150)$  \\
        Laiho 11      & \cite{Laiho:2011np} & $0.7628(38)(205)$  \\
        BMW 11        & \cite{Durr:2011ap}  & $0.7727(81)(84)$  \\ \hline
        FLAG-23       & \cite{FlavourLatticeAveragingGroupFLAG:2021npn}
                                            & $0.7625(97)$
        \\ \hline\hline
      \end{tabular}
    } 
    \caption{$\BK$}
    \label{tab:BK}
  \end{subtable} 
  \hfill
  \begin{subtable}{0.57\linewidth}
    \renewcommand{\arraystretch}{1.2}
    \resizebox{1.0\linewidth}{!}{
      \begin{tabular}{ @{\qquad} c @{\qquad} l @{\qquad} l @{\qquad} }
        \hline\hline
        Input & Value & Ref. \\ \hline
        $G_{F}$
        & \red{ $1.1663788(6) \times 10^{-5}$ GeV$^{-2}$ }
        & PDG-23 \cite{ Workman:2022ynf} \\ \hline
        $\theta$
        & $43.52(5)^{\circ}$
        & PDG-23 \cite{ Workman:2022ynf} \\ \hline
        $m_{K^{0}}$
        & $497.611(13)$ MeV
        & PDG-23 \cite{ Workman:2022ynf} \\ \hline
        $\Delta M_{K}$
        & $3.484(6) \times 10^{-12}$ MeV
        & PDG-23 \cite{ Workman:2022ynf} \\ \hline
        $F_K$
        & $155.7(3)$ MeV
        & FLAG-23 \cite{ FlavourLatticeAveragingGroupFLAG:2021npn}
        \\ \hline\hline
      \end{tabular}
    } 
    \caption{Other parameters}
    \label{tab:other}
  \end{subtable} 
  \caption{ (\subref{tab:BK}) Results for $\BK$ and
    (\subref{tab:other}) other input parameters.}
  \label{tab:input-BK-other}
\end{table}

The dispersive long distance (LD) effect $\xi_\text{LD}$ is 
\begin{align}
  \xi_\text{LD} &=  \frac{m^\prime_\text{LD}}{\sqrt{2} \Delta M_K} \,,
  \qquad
  m^\prime_\text{LD}
  = -\Im \left[ \mathcal{P}\sum_{C}
    \frac{\mate{\wbar{K}^0}{H_\text{w}}{C} \mate{C}{H_\text{w}}{K^0}}
         {m_{K^0}-E_{C}}  \right]
  \label{eq:xi-LD}
\end{align}
As explained in Ref.~\cite{ Bailey:2018feb}, there are two independent
methods to estimate $\xi_\text{LD}$: one is the BGI estimate \cite{
  Buras:2010}, and the other is the RBC-UKQCD estimate \cite{
  Christ:2012, Christ:2014qwa}.
The BGI method estimates $\xi_\text{LD}$ using chiral perturbation
theory, using Eq.~\eqref{eq:xiLD:bgi}.
\begin{align}
  \xi_\text{LD} &= -0.4(3) \times \frac{\xi_0}{ \sqrt{2} }
  \label{eq:xiLD:bgi}
\end{align}
The RBC-UKQCD method estimates $\xi_\text{LD}$ using
Eq.~\eqref{eq:xiLD:rbc}.
\begin{align}
  \xi_\text{LD} &= (0 \pm 1.6)\%.
  \label{eq:xiLD:rbc}
\end{align}
We use both methods to estimate the size of $\xi_\text{LD}$.

In Table \ref{tab:input-WP-eta} (\subref{tab:eta}), we present higher
order QCD corrections: $\eta_{ij}$ with $i,j = t,c$.
A new approach using $u-t$ unitarity instead of $c-t$ unitarity
appeared in Ref.~\cite{ Brod:2019rzc}, which is supposed to have 
better convergence with respect to the charm quark mass.
We are working to incorporate this into our analysis, which we will
report soon.

In Table \ref{tab:input-BK-other} (\subref{tab:other}), we present other
input parameters needed to evaluate $\epsK$.
Note that the Fermi coupling constant $G_F$ has been updated in 2023.


%
%
\section{Quark masses}
In Table \ref{tab:m_c:m_t} we present the charm quark mass $m_c(m_c)$
and top quark mass $m_t(m_t)$.
From FLAG 2023 \cite{ FlavourLatticeAveragingGroupFLAG:2021npn}, we
take the results for $m_c (m_c)$ with $N_f = 2+1$, since there is some
inconsistency among the lattice results of various groups with $N_f =
2+1+1$.
For the top quark mass, we use the PDG 2023 results for the pole mass
$M_t$ to obtain $m_t (m_t)$\footnote{Here we use PDG results updated
on 2023-5-31.}.
\begin{table}[t!]
  \begin{subtable}{0.48\linewidth}
    \vspace*{-5mm}
    \renewcommand{\arraystretch}{1.6}
    \resizebox{0.99\linewidth}{!}{
      \begin{tabular}{ l @{\qquad} l @{\qquad} l @{\qquad} l }
        \hline\hline
        {\small Collaboration} & $N_f$ & $m_c(m_c)$ & Ref.
        \\ \hline
        FLAG 2023       & $2+1$   & \red{1.276(5)}
        & \cite{FlavourLatticeAveragingGroupFLAG:2021npn}
        \\
        FLAG 2023       & $2+1+1$ & \blue{1.280(13)}
        & \cite{FlavourLatticeAveragingGroupFLAG:2021npn}
        \\ \hline\hline
      \end{tabular}
    } 
    \caption{$m_c(m_c)$ [GeV]}
    \label{tab:m_c}
  \end{subtable} 
  \hfill
  \begin{subtable}{0.50\linewidth}
    \renewcommand{\arraystretch}{1.2}
    \vspace*{-5mm}
    \resizebox{0.99\linewidth}{!}{
      \begin{tabular}{ l l l l }
        \hline\hline
        {\small Collaboration} & $M_t$ & $m_t(m_t)$ & Ref.
        \\ \hline
        PDG 2021 & 172.76(30) & 162.96(28)(17) & \cite{Zyla:2020zbs}
        \\
        PDG 2022 & 172.69(30) & 162.90(28)(17) & \cite{Workman:2022ynf}
        \\
        PDG 2023 & 172.69(30) & \red{ 162.90(28)(17) } & \cite{Workman:2022ynf}
        \\ \hline\hline
      \end{tabular}
    } 
    \caption{ $m_t(m_t)$ [GeV] }
    \label{tab:m_t}
  \end{subtable} 
  \caption{  Results for (\subref{tab:m_c}) charm quark mass and
    (\subref{tab:m_t}) top quark mass. }
  \label{tab:m_c:m_t}
\end{table}

In Table \ref{tab:M_t+err_bud}\;(\subref{fig:M_t}) we present the
history of values for the top quark pole mass $M_t$.
We find that the average value shifts downward by 0.47\% over time.
We also find that the error shrinks fast down to 30\% of the original
error (2012) thanks to accumulation of high statistics in the LHC
experiments.
The data for 2020 is dropped out intentionally in memory of the
absence of Lattice 2020 due to COVID-19.
\begin{table}[h!]
  \begin{subfigure}{0.53\linewidth}
    \vspace{-7mm}
    \includegraphics[width=1.0\linewidth]{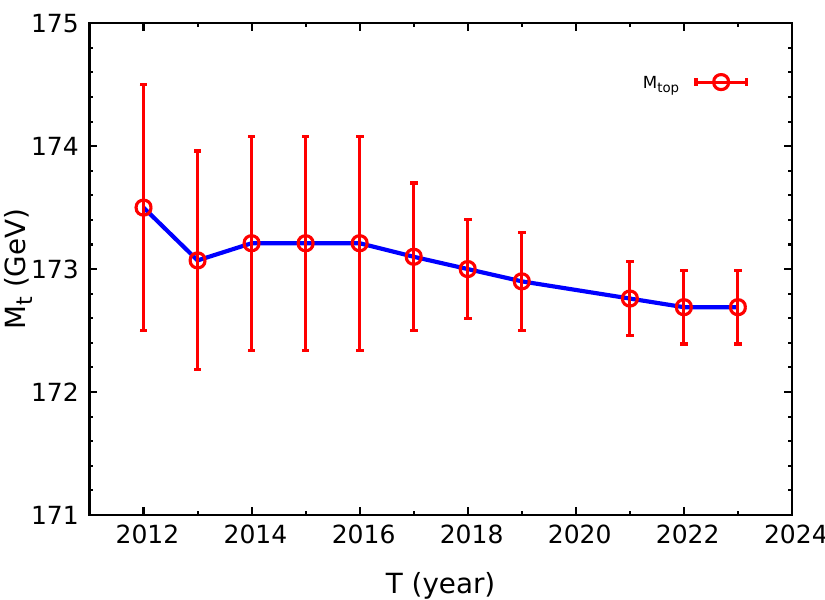}
    \caption{History of $M_t$ (top quark pole mass).}
    \label{fig:M_t}
  \end{subfigure}
  \hfill
  \begin{subtable}{0.44\linewidth}
    \vspace{-7mm}
    \renewcommand{\arraystretch}{1.17}
    \resizebox{0.99\linewidth}{!}{
    \begin{tabular}{@{\qquad} l @{\qquad} l @{\qquad} l }
      \hline\hline
      source & error (\%) & memo 
      \\ \hline
      $\Vcb$          & \red{ 49.6 }     & Exclusive \\
      $\eta_{ct}$     & 20.7             & $c-t$ Box \\        
      $\bar\eta$      & 13.3             & AOF \\
      $\eta_{cc}$     & \phantom{0}8.8   & $c-c$ Box \\        
      $\xi_\text{LD}$ & \phantom{0}2.1   & RBC-UKQCD \\        
      $\bar\rho$      & \phantom{0}2.1   & AOF \\        
      $\BK$           & \phantom{0}1.7   & FLAG \\
      $\;\vdots$      & $\;\;\;\;\vdots$ & $\;\;\vdots$
      \\ \hline\hline
    \end{tabular}
    } 
    \vspace*{4mm}
    \caption{Error budget for $|\epsK|^\text{SM}$ }
    \label{tab:err-bud}
  \end{subtable}
  \caption{(\subref{fig:M_t}) $M_t$ history (\subref{tab:err-bud})
    error budget for $|\epsK|^\text{SM}$. }
  \label{tab:M_t+err_bud}
\end{table}

%

%
%

%
%
\section{$W$ boson mass}
\begin{table}[t!]
  \begin{subfigure}{0.53\linewidth}
    \vspace{-3mm}
    \includegraphics[width=1.0\linewidth]{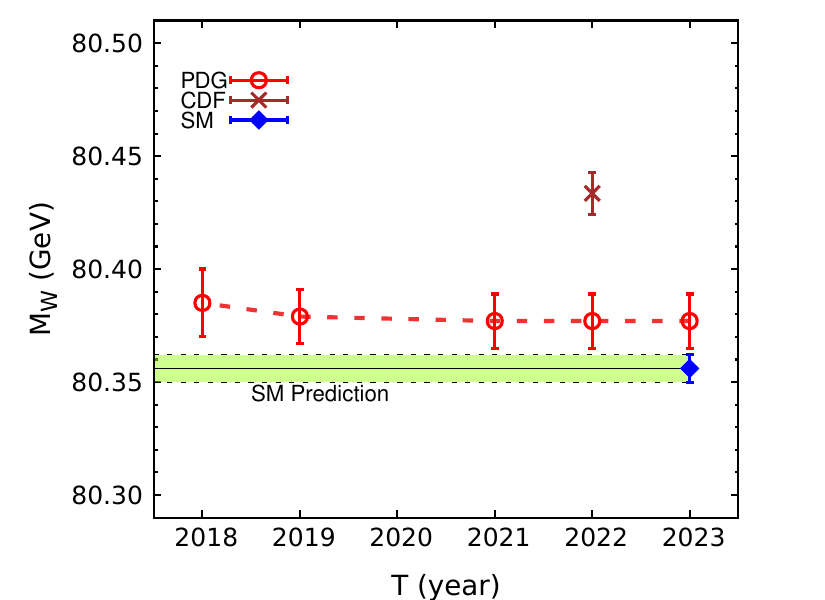}
    \caption{History of $M_W$.}
    \label{fig:m_W}
  \end{subfigure}
  \hfill
  \begin{subtable}{0.45\linewidth}
    \vspace{-3mm}
    \renewcommand{\arraystretch}{1.35}
    \resizebox{0.99\linewidth}{!}{
    \begin{tabular}{@{\qquad} l @{\qquad} l @{\qquad} l @{\qquad} }
      \hline\hline
      Source & $M_W$ (GeV) & Ref. 
      \\ \hline
      SM-2023  & \red{ 80.356(6) }   & \cite{ Workman:2022ynf}   \\
      CDF-2022 &       80.4335(94)   & \cite{ CDF:2022hxs}       \\
      PDG-2023 &       80.377(12)    & \cite{ Workman:2022ynf}   \\
      PDG-2022 &       80.377(12)    & \cite{ Workman:2022ynf}   \\
      PDG-2021 &       80.377(12)    & \cite{ Zyla:2020zbs}      \\
      PDG-2019 &       80.379(12)    & \cite{ Tanabashi:2018oca} \\
      PDG-2018 &       80.385(15)    & \cite{ Patrignani:2016xqp}
      \\ \hline\hline
    \end{tabular}
    } 
    \vspace*{4mm}
    \caption{Table of $M_W$. }
    \label{tab:m_W}
  \end{subtable}
  \caption{$W$-boson mass: (\subref{fig:m_W}) $M_W$ versus time,
    and (\subref{tab:m_W}) table of $M_W$. }
  \label{tab:m_W+fig}
\end{table}

In Fig.~\ref{tab:m_W+fig}\;(\subref{fig:m_W}) we plot $M_W$ (the $W$
boson mass) as a function of time.
The corresponding results for $M_W$ are summarized in Table
\ref{tab:m_W+fig}\;(\subref{tab:m_W}).
In Fig.~\ref{tab:m_W+fig}\;(\subref{fig:m_W}), the light-green band
represents the standard model (SM) prediction, the red circles
represent the PDG results from the experimental summary, and the brown
cross represents the CDF-2022 result.
The upside is that the CDF-2022 result is the most precise
experimental result for $M_W$.
The downside, however, is that it deviates by $6.9\sigma$ from the
standard model prediction (SM-2023).
Here we use the SM-2023 result for $M_W$ to evaluate $\epsK$.
%
%

%
%
\section{Results for $\epsK$}
\begin{figure}[t!]
  \begin{subfigure}{0.47\linewidth}
    \vspace*{-5mm}
    \includegraphics[width=1.0\linewidth]
       {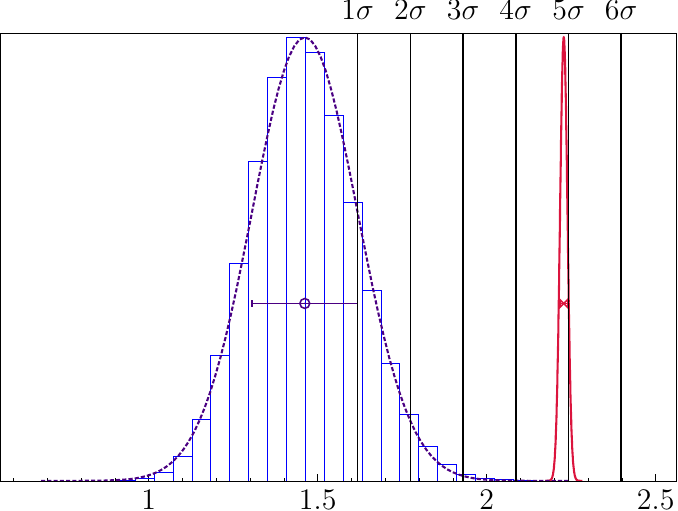}
    \caption{Exclusive $\Vcb$ (FNAL/MILC-22, BGL)}
    \label{fig:epsK-ex:rbc}
  \end{subfigure}
  \hfill
  \begin{subfigure}{0.47\linewidth}
    \vspace*{-5mm}
    \includegraphics[width=1.0\linewidth]
                    {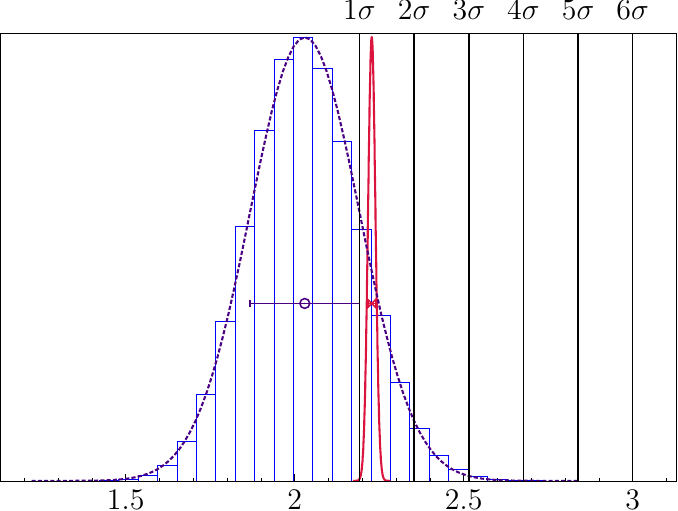}
    \caption{Inclusive $\Vcb$ (HFLAV-23, 1S scheme)}
    \label{fig:epsK-in:rbc}
  \end{subfigure}
  \caption{$|\epsK|$ with (\subref{fig:epsK-ex:rbc}) exclusive $\Vcb$
    (left) and (\subref{fig:epsK-in:rbc}) inclusive $\Vcb$ (right) in
    units of $1.0\times 10^{-3}$. }
  \label{fig:epsK:cmp:rbc}
\end{figure}

In Fig.~\ref{fig:epsK:cmp:rbc} we show results for $|\epsK|$ evaluated
directly from the standard model (SM) with the lattice QCD inputs
given in the previous sections.
In Fig.~\ref{fig:epsK:cmp:rbc}\;(\subref{fig:epsK-ex:rbc}), the blue
curve represents the theoretical evaluation of $|\epsK|$ obtained
using the FLAG-23 results for $\BK$, AOF for Wolfenstein parameters,
the FNAL/MILC-22 results for exclusive $\Vcb$, results for $\xi_0$
with the indirect method, and the RBC-UKQCD estimate for
$\xi_\text{LD}$.
The red curve in Fig.~\ref{fig:epsK:cmp:rbc} represents the experimental
results for $|\epsK|$.
In Fig.~\ref{fig:epsK:cmp:rbc}\;(\subref{fig:epsK-in:rbc}), the blue
curve represents the theoretical evaluation of $|\epsK|$ obtained
using the same input parameters as in
Fig.~\ref{fig:epsK:cmp:rbc}\;(\subref{fig:epsK-ex:rbc})
except for $\Vcb$.
Here we use the 1S-scheme results for inclusive $\Vcb$ instead of
those for exclusive $\Vcb$.

Our results for $|\epsK|^\text{SM}$ and $\Delta\epsK =
|\epsK^\text{Exp}| - |\epsK^\text{SM}|$ are summarized in Table
\ref{tab:epsK}.
Here, the superscript ${}^\text{SM}$ represents the theoretical
expectation value of $|\epsK|$ obtained directly from the SM.
The superscript ${}^\text{Exp}$ represents the experimental value
of $|\epsK| = 2.228(11) \times 10^{-3}$.
Results in Table \ref{tab:epsK}\;(\subref{tab:epsK:rbc}) are obtained
using the RBC-UKQCD estimate for $\xi_\text{LD}$, and those in
Table \ref{tab:epsK}\;(\subref{tab:epsK:bgi}) are obtained using
the BGI estimate for $\xi_\text{LD}$.
In Table \ref{tab:epsK}\;(\subref{tab:epsK:rbc}), we find that the
theoretical expectation values of $|\epsK|^\text{SM}$ with lattice QCD
inputs (with exclusive $\Vcb$) have $4.9\sigma \sim 3.9\sigma$ tension
with the experimental value of $|\epsK|^\text{Exp}$, while there is no
tension with inclusive $\Vcb$ (obtained using the heavy quark
expansion and QCD sum rules).

\begin{table}[b!]
%
  \begin{subtable}{1.0\linewidth}
    \center
    \renewcommand{\arraystretch}{1.2}
    \resizebox{0.85\linewidth}{!}{
      \begin{tabular}{@{\qquad} l @{\qquad} l @{\qquad} l @{\qquad} l @{\qquad} l @{\qquad} }
        \hline\hline
        $\Vcb$    & method   & source    & $|\epsK|^\text{SM}$ & $\Delta\epsK$
        \\ \hline
        exclusive & BGL      & FNAL/MILC-22 & $1.462 \pm 0.156$ & $4.90\sigma$
        \\
        exclusive & comb     & HFLAV-23     & $1.561 \pm 0.138$ & $4.84\sigma$
        \\        
        exclusive & comb     & FLAG-23      & $1.619 \pm 0.157$ & $3.88\sigma$
        \\
        exclusive & comb     & HPQCD-23     & $1.594 \pm 0.162$ & $3.90\sigma$
        \\ \hline
        inclusive & 1S       & HFLAV-23     & $2.030 \pm 0.162$ & $1.22\sigma$
        \\
        inclusive & kinetic  & Gambino-21   & $2.063 \pm 0.169$ & $0.98\sigma$
        \\ \hline\hline
      \end{tabular}
    } 
    \caption{$|\epsK|$ with RBC-UKQCD estimate for $\xi_\text{LD}$}
    \label{tab:epsK:rbc}
  \end{subtable} 
  \begin{subtable}{1.0\linewidth}
    \vspace*{3mm}
    \center
    \renewcommand{\arraystretch}{1.2}
    \resizebox{0.85\linewidth}{!}{
      \begin{tabular}{@{\qquad} l @{\qquad} l @{\qquad} l @{\qquad} l @{\qquad} l @{\qquad} }
        \hline\hline
        $\Vcb$    & method   & reference  & $|\epsK|^\text{SM}$ & $\Delta\epsK$
        \\ \hline
        exclusive & BGL  & FNAL/MILC-21 & $1.510 \pm 0.159$ & $4.52\sigma$
        \\
        exclusive & comb & HFLAV-23     & $1.609 \pm 0.140$ & $4.41\sigma$
        \\ \hline\hline
      \end{tabular}
    } 
    \caption{$|\epsK|$ with BGI estimate for $\xi_\text{LD}$}
    \label{tab:epsK:bgi}
  \end{subtable} 
  \caption{ $|\epsK|$ in units of $1.0\times 10^{-3}$, and
    $\Delta\epsK = |\epsK|^\text{Exp} - |\epsK|^\text{SM}$ in units of
    $\sigma$.}
  \label{tab:epsK}
\end{table}

\begin{figure}[t!]
  \begin{subfigure}{0.491\linewidth}
    \vspace*{-6mm}
    \includegraphics[width=\linewidth]
                    {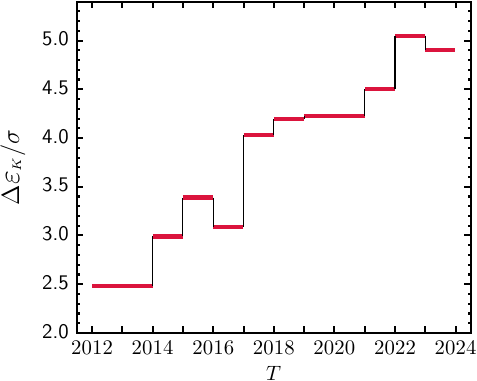}
    \caption{Time evolution of $\Delta \epsK/\sigma$}
    \label{fig:depsK:rbc:his}
  \end{subfigure}
  \hfill
  \begin{subfigure}{0.469\linewidth}
    \vspace*{-6mm}
    \includegraphics[width=\linewidth]
                    {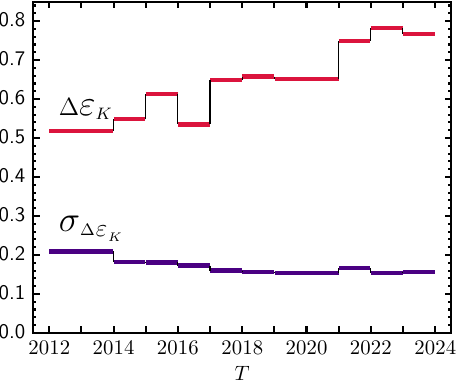}
    \caption{Time evolution of the average and error of $\Delta\epsK$}
    \label{fig:depsK+sigma:rbc:his}
  \end{subfigure}
  \caption{ Time history of (\subref{fig:depsK:rbc:his})
    $\Delta\epsK/\sigma$, and (\subref{fig:depsK+sigma:rbc:his})
    $\Delta\epsK$ and $\sigma_{\Delta\epsK}$. }
  \label{fig:depsK:sum:rbc:his}
\end{figure}

In Fig.~\ref{fig:depsK:sum:rbc:his}\;(\subref{fig:depsK:rbc:his}), we
show the time evolution of $\Delta\epsK/\sigma$ starting from 2012
till 2023.
In 2012, $\Delta\epsK$ was $2.5\sigma$, but now it is $4.9\sigma$ with
exclusive $\Vcb$ (FNAL/MILC-22, BGL).
Here we use the results for exclusive $\Vcb$ from FNAL/MILC-22, since
it contains the most comprehensive analysis of the $\BtoDst$ decays at
both zero recoil and non-zero recoil, while it incorporates both BELLE
and BABAR experimental results.
In Fig.~\ref{fig:depsK:sum:rbc:his} (\subref{fig:depsK+sigma:rbc:his})
we show the time evolution of the average $\Delta\epsK$ and the error
$\sigma_{\Delta\epsK}$ during the period of 2012--2023.

At present we find $\Vcb$ gives the largest error ($\approx 50\%$) in
$|\epsK|^\text{SM}$.
Refer to Table \ref{tab:M_t+err_bud} (\subref{tab:err-bud}) for more
details.
Hence, it is essential to reduce the errors in $\Vcb$ as much as
possible.
Part of the errors come from experiments in BELLE, BELLE2, BABAR, and
LHCb, which are beyond our control but will decrease thanks to
on-going accumulation of higher statistics in BELLE2 and LHCb.
Part of the errors come from the theory used to evaluate the
semi-leptonic form factors for $\BtoDstp$ decays, using the tools of
lattice QCD.
In order to reduce the errors on the theoretical side, there is an
on-going project to determine exclusive $\Vcb$ using the
Oktay-Kronfeld (OK) action for the heavy quarks to calculate the form
factors for $\BtoDstp$ decays \cite{ Bhattacharya:2021peq,
  Park:2020vso, Bhattacharya:2020xyb, Bhattacharya:2018ibo,
  Bailey:2017xjk, Bailey:2017zgt, Bailey:2020uon}.

A large portion of interesting results for $|\epsK|^\text{SM}$ and
$\Delta\epsK$ could not be presented in Table \ref{tab:epsK} and in
Fig.~\ref{fig:depsK:sum:rbc:his} due to lack of space: For example,
results for $|\epsK|^\text{SM}$ obtained using exclusive $\Vcb$
(FLAG-23) with the BGI estimate for $\xi_\text{LD}$, results for
$|\epsK|^\text{SM}$ obtained using $\xi_0$ determined by the direct
method, and so on.
We plan to report them collectively in Ref.~\cite{ wlee:2023epsK}.
We find that there was another analysis on $\epsK$ in Ref.~\cite{
  Buras:2022wpw}.


%
%

\acknowledgments
We thank J.~Bailey, Y.C.~Jang, S.~Sharpe, and R.~Gupta for helpful
discussion.
We thank G.~Martinelli for providing us with the most updated results
of the UTfit Collaboration in time.
The research of W.~Lee is supported by the Mid-Career Research
Program (Grant No.~NRF-2019R1A2C2085685) of the NRF grant funded by
the Korean government (MSIT).
W.~Lee would like to acknowledge the support from the KISTI
supercomputing center through the strategic support program for the
supercomputing application research (No.~KSC-2018-CHA-0043,
KSC-2020-CHA-0001, KSC-2023-CHA-0010).
Computations were carried out in part on the DAVID cluster at Seoul
National University.
%

\bibliography{refs}

\providecommand{\href}[2]{#2}\begingroup\raggedright\begin{thebibliography}{10}

\bibitem{Lee:2023lxz}
{\bf SWME} Collaboration, W.~Lee, S.~Kim, S.~Lee, J.~Leem, and S.~Park, {\it
  {2022 update on $\varepsilon_K$ with lattice QCD inputs}},  {\em PoS} {\bf
  LATTICE2022} (2023) 297, [\href{http://xxx.lanl.gov/abs/2301.12375}{{\tt
  2301.12375}}].

\bibitem{Lee:2021crz}
{\bf SWME} Collaboration, W.~Lee, J.~Kim, Y.-C. Jang, S.~Lee, J.~Leem, C.~Park,
  and S.~Park, {\it {2021 update on $\varepsilon_K$ with lattice QCD inputs}},
  {\em PoS} {\bf LATTICE2021} (2021) 078,
  [\href{http://xxx.lanl.gov/abs/2202.11473}{{\tt 2202.11473}}].

\bibitem{Kim:2019vic}
{\bf LANL-SWME} Collaboration, J.~Kim, S.~Lee, W.~Lee, Y.-C. Jang, J.~Leem, and
  S.~Park {\em PoS} {\bf LATTICE2019} (2019) 029,
  [\href{http://xxx.lanl.gov/abs/1912.03024}{{\tt 1912.03024}}].

\bibitem{Bailey:2018feb}
J.~A. Bailey {\em et~al.} {\em Phys. Rev.} {\bf D98} (2018) 094505,
  [\href{http://xxx.lanl.gov/abs/1808.09657}{{\tt 1808.09657}}].

\bibitem{Bailey:2015tba}
J.~A. Bailey, Y.-C. Jang, W.~Lee, and S.~Park {\em Phys. Rev.} {\bf D92} (2015)
  034510, [\href{http://xxx.lanl.gov/abs/1503.05388}{{\tt 1503.05388}}].

\bibitem{Bailey:2018aks}
J.~A. Bailey {\em et~al.} {\em PoS} {\bf LATTICE2018} (2018) 284,
  [\href{http://xxx.lanl.gov/abs/1810.09761}{{\tt 1810.09761}}].

\bibitem{Jang:2017ieg}
Y.-C. Jang, W.~Lee, S.~Lee, and J.~Leem {\em EPJ Web Conf.} {\bf 175} (2018)
  14015, [\href{http://xxx.lanl.gov/abs/1710.06614}{{\tt 1710.06614}}].

\bibitem{Bailey:2015frw}
J.~A. Bailey, Y.-C. Jang, W.~Lee, and S.~Park {\em PoS} {\bf LATTICE2015}
  (2015) 348, [\href{http://xxx.lanl.gov/abs/1511.00969}{{\tt 1511.00969}}].

\bibitem{FlavourLatticeAveragingGroupFLAG:2021npn}
{\bf Flavour Lattice Averaging Group (FLAG)} Collaboration, Y.~Aoki {\em
  et~al.} {\em Eur. Phys. J. C} {\bf 82} (2022), no.~10 869,
  [\href{http://xxx.lanl.gov/abs/2111.09849}{{\tt 2111.09849}}].

\bibitem{Zyla:2020zbs}
{\bf Particle Data Group} Collaboration, P.~Zyla {\em et~al.} {\em PTEP} {\bf
  2020} (2020), no.~8 083C01.

\bibitem{UTfit:2022hsi}
{\bf UTfit} Collaboration, M.~Bona {\em et~al.}, {\it {New UTfit Analysis of
  the Unitarity Triangle in the Cabibbo-Kobayashi-Maskawa scheme}},  {\em Rend.
  Lincei Sci. Fis. Nat.} {\bf 34} (2023) 37--57,
  [\href{http://xxx.lanl.gov/abs/2212.03894}{{\tt 2212.03894}}].

\bibitem{Charles:2004jd}
J.~Charles {\em et~al.} {\em Eur.Phys.J.} {\bf C41} (2005) 1--131,
  [\href{http://xxx.lanl.gov/abs/hep-ph/0406184}{{\tt hep-ph/0406184}}].
  updated results and plots available at: \url{http://ckmfitter.in2p3.fr}.

\bibitem{Bona:2006ah}
M.~Bona {\em et~al.} {\em JHEP} {\bf 10} (2006) 081,
  [\href{http://xxx.lanl.gov/abs/hep-ph/0606167}{{\tt hep-ph/0606167}}].
  {Standard Model fit results: Summer 2016 (ICHEP 2016):
  \url{http://www.utfit.org}}.

\bibitem{Buras2008:PhysRevD.78.033005}
A.~J. Buras and D.~Guadagnoli {\em Phys.Rev.} {\bf D78} (2008) 033005,
  [\href{http://xxx.lanl.gov/abs/0805.3887}{{\tt 0805.3887}}].

\bibitem{Brod2010:prd.82.094026}
J.~Brod and M.~Gorbahn {\em Phys.Rev.} {\bf D82} (2010) 094026,
  [\href{http://xxx.lanl.gov/abs/1007.0684}{{\tt 1007.0684}}].

\bibitem{Barone:2022gkn}
A.~Barone, A.~J\"uttner, S.~Hashimoto, T.~Kaneko, and R.~Kellermann, {\it
  {Inclusive semi-leptonic $B_{(s)}$ mesons decay at the physical $b$ quark
  mass}},  {\em PoS} {\bf LATTICE2022} (2023) 403,
  [\href{http://xxx.lanl.gov/abs/2211.15623}{{\tt 2211.15623}}].

\bibitem{FermilabLattice:2021cdg}
{\bf Fermilab Lattice, MILC} Collaboration, A.~Bazavov {\em et~al.} {\em Eur.
  Phys. J. C} {\bf 82} (2022), no.~12 1141,
  [\href{http://xxx.lanl.gov/abs/2105.14019}{{\tt 2105.14019}}].

\bibitem{HeavyFlavorAveragingGroup:2022wzx}
{\bf Heavy Flavor Averaging Group, HFLAV} Collaboration, Y.~S. Amhis {\em
  et~al.}, {\it {Averages of b-hadron, c-hadron, and \ensuremath{\tau}-lepton
  properties as of 2021}},  {\em Phys. Rev. D} {\bf 107} (2023), no.~5 052008,
  [\href{http://xxx.lanl.gov/abs/2206.07501}{{\tt 2206.07501}}].

\bibitem{Harrison:2023dzh}
J.~Harrison and C.~T.~H. Davies, {\it {$B \rightarrow D^*$ vector, axial-vector
  and tensor form factors for the full $q^2$ range from lattice QCD}},
  \href{http://xxx.lanl.gov/abs/2304.03137}{{\tt 2304.03137}}.

\bibitem{Bordone:2021oof}
M.~Bordone, B.~Capdevila, and P.~Gambino {\em Phys. Lett. B} {\bf 822} (2021)
  136679, [\href{http://xxx.lanl.gov/abs/2107.00604}{{\tt 2107.00604}}].

\bibitem{Blum:2015ywa}
T.~Blum {\em et~al.} {\em Phys. Rev.} {\bf D91} (2015) 074502,
  [\href{http://xxx.lanl.gov/abs/1502.00263}{{\tt 1502.00263}}].

\bibitem{Bai:2015nea}
Z.~Bai {\em et~al.} {\em Phys. Rev. Lett.} {\bf 115} (2015) 212001,
  [\href{http://xxx.lanl.gov/abs/1505.07863}{{\tt 1505.07863}}].

\bibitem{RBC:2020kdj}
{\bf RBC, UKQCD} Collaboration, R.~Abbott {\em et~al.} {\em Phys. Rev. D} {\bf
  102} (2020), no.~5 054509, [\href{http://xxx.lanl.gov/abs/2004.09440}{{\tt
  2004.09440}}].

\bibitem{Jang:2015sla}
B.~J. Choi {\em et~al.} {\em Phys. Rev.} {\bf D93} (2016) 014511,
  [\href{http://xxx.lanl.gov/abs/1509.00592}{{\tt 1509.00592}}].

\bibitem{Blum:2014tka}
T.~Blum {\em et~al.} {\em Phys. Rev.} {\bf D93} (2016) 074505,
  [\href{http://xxx.lanl.gov/abs/1411.7017}{{\tt 1411.7017}}].

\bibitem{Laiho:2011np}
J.~Laiho and R.~S. Van~de Water {\em PoS} {\bf LATTICE2011} (2011) 293,
  [\href{http://xxx.lanl.gov/abs/1112.4861}{{\tt 1112.4861}}].

\bibitem{Durr:2011ap}
S.~Durr {\em et~al.} {\em Phys. Lett.} {\bf B705} (2011) 477--481,
  [\href{http://xxx.lanl.gov/abs/1106.3230}{{\tt 1106.3230}}].

\bibitem{Workman:2022ynf}
{\bf Particle Data Group} Collaboration, R.~L. Workman and Others {\em PTEP}
  {\bf 2022} (2022) 083C01.

\bibitem{Buras:2010}
A.~J. Buras, D.~Guadagnoli, and G.~Isidori {\em Phys.Lett.} {\bf B688} (2010)
  309--313, [\href{http://xxx.lanl.gov/abs/1002.3612}{{\tt 1002.3612}}].

\bibitem{Christ:2012}
N.~Christ {\em et~al.} {\em Phys.Rev.} {\bf D88} (2013) 014508,
  [\href{http://xxx.lanl.gov/abs/1212.5931}{{\tt 1212.5931}}].

\bibitem{Christ:2014qwa}
N.~Christ {\em et~al.} {\em PoS} {\bf LATTICE2013} (2014) 397,
  [\href{http://xxx.lanl.gov/abs/1402.2577}{{\tt 1402.2577}}].

\bibitem{Brod:2019rzc}
J.~Brod, M.~Gorbahn, and E.~Stamou {\em Phys. Rev. Lett.} {\bf 125} (2020),
  no.~17 171803, [\href{http://xxx.lanl.gov/abs/1911.06822}{{\tt 1911.06822}}].

\bibitem{CDF:2022hxs}
{\bf CDF} Collaboration, T.~Aaltonen {\em et~al.} {\em Science} {\bf 376}
  (2022), no.~6589 170--176.

\bibitem{Tanabashi:2018oca}
M.~Tanabashi {\em et~al.} {\em Phys. Rev.} {\bf D98} (2018) 030001.
  \url{http://pdg.lbl.gov/2019/}.

\bibitem{Patrignani:2016xqp}
C.~Patrignani {\em et~al.} {\em Chin. Phys.} {\bf C40} (2016) 100001.
  {\url{https://pdg.lbl.gov/} }.

\bibitem{Bhattacharya:2021peq}
T.~Bhattacharya, B.~J. Choi, R.~Gupta, Y.-C. Jang, S.~Jwa, S.~Lee, W.~Lee,
  J.~Leem, S.~Park, and B.~Yoon {\em PoS} {\bf LATTICE2021} (2021) 136,
  [\href{http://xxx.lanl.gov/abs/2204.05848}{{\tt 2204.05848}}].

\bibitem{Park:2020vso}
{\bf LANL-SWME} Collaboration, S.~Park, T.~Bhattacharya, R.~Gupta, Y.-C. Jang,
  B.~J. Choi, S.~Jwa, S.~Lee, W.~Lee, and J.~Leem {\em PoS} {\bf LATTICE2019}
  (2020) 050, [\href{http://xxx.lanl.gov/abs/2002.04755}{{\tt 2002.04755}}].

\bibitem{Bhattacharya:2020xyb}
{\bf LANL/SWME} Collaboration, T.~Bhattacharya, B.~J. Choi, R.~Gupta, Y.-C.
  Jang, S.~Jwa, S.~Lee, W.~Lee, J.~Leem, and S.~Park {\em PoS} {\bf
  LATTICE2019} (2020) 056, [\href{http://xxx.lanl.gov/abs/2003.09206}{{\tt
  2003.09206}}].

\bibitem{Bhattacharya:2018ibo}
T.~Bhattacharya {\em et~al.} {\em PoS} {\bf LATTICE2018} (2018) 283,
  [\href{http://xxx.lanl.gov/abs/1812.07675}{{\tt 1812.07675}}].

\bibitem{Bailey:2017xjk}
J.~A. Bailey {\em et~al.} {\em EPJ Web Conf.} {\bf 175} (2018) 13012,
  [\href{http://xxx.lanl.gov/abs/1711.01786}{{\tt 1711.01786}}].

\bibitem{Bailey:2017zgt}
J.~Bailey, Y.-C. Jang, W.~Lee, and J.~Leem {\em EPJ Web Conf.} {\bf 175} (2018)
  14010, [\href{http://xxx.lanl.gov/abs/1711.01777}{{\tt 1711.01777}}].

\bibitem{Bailey:2020uon}
{\bf LANL-SWME} Collaboration, J.~A. Bailey, Y.-C. Jang, S.~Lee, W.~Lee, and
  J.~Leem {\em Phys. Rev. D} {\bf 105} (2022), no.~3 034509,
  [\href{http://xxx.lanl.gov/abs/2001.05590}{{\tt 2001.05590}}].

\bibitem{wlee:2023epsK}
{\bf SWME} Collaboration, J.~Bailey, J.~Kim, S.~Lee, W.~Lee, Y.-C. Jang,
  J.~Leem, S.~Park, {\em et~al.}
\newblock {in preparation}.

\bibitem{Buras:2022wpw}
A.~J. Buras and E.~Venturini, {\it {The exclusive vision of rare K and B decays
  and of the quark mixing in the standard model}},  {\em Eur. Phys. J. C} {\bf
  82} (2022), no.~7 615, [\href{http://xxx.lanl.gov/abs/2203.11960}{{\tt
  2203.11960}}].

\end{thebibliography}\endgroup


\end{document}